\documentclass[11pt]{article}
\usepackage{amsmath}
\usepackage{epsfig,bm}
\def\b{\begin}
\def\e{\end}
\def\t{thebibliography}
\def\bi{\bibitem}
\def\be{\b{equation}}
\def\ee{\e{equation}}

\def\nt{\noindent}
\def\tr{\mbox{Tr}}

\begin{document}

\title{A GENERAL INFORMATION THEORETICAL PROOF FOR THE SECOND LAW OF
THERMODYNAMICS }

\author{Qi-Ren Zhang
\\Department of Technical Physics, Peking University , Beijing,100871,
China}

\maketitle

\vskip0.3cm

\begin{abstract}
We show that the conservation and the non-additivity of the
information, together with the additivity of the entropy make the
entropy increase in an isolated system.  The collapse of the
entangled quantum state offers an example of the information
non-additivity. Nevertheless, the later is also true in other
fields, in which the interaction information is important. Examples
are classical statistical mechanics, social statistics and financial
processes. The second law of thermodynamics is thus proven in its
most general form. It is exactly true, not only in quantum and
classical physics but also in other processes, in which the
information is conservative and non-additive.
\end{abstract}

\bigskip

To understand the foundation of the second law of thermodynamics is
a long standing problem in physics. Text books tell us\cite{l}, the
state of a macroscopic system with larger entropy is more probable.
However, whether the system always goes from a less probable state
to a more probable state is still an open question. The H-theorem of
Boltzmann is a classical proof for definite approaching to
equilibrium. It is based on a model of colliding classical particle
system for the macroscopic matter, therefore is not general enough,
even from the view point of the classical statistical physics. In
1948, Shannon\cite{s1,s2} discovered a powerful theory of
information. It is applicable in analysis of all statistical
processes, including statistical physics. Here we show that by use
of its fundamental ideas, one can simply prove the second law of
thermodynamics in its most general form. In quantum statistical
mechanics, it is based on the state entanglement in time development
and the state collapse in measurement, therefore is quite general.

The time development of the density operator $\rho(t)$ for an
isolated system is governed by the von Neumann equation. Its
solution is \be \rho(t)=U(t,t_0)\rho(t_0)U(t_0,t)\; ,\label{1}\ee in
which $U(t,t_0)$ is the time displacement operator of the state from
time $t_0$ to time $t$. Defining the information \be {\cal I}(t)=\tr
[\rho(t)\ln\rho(t)]\label{2}\ee at time t, we see from (\ref{1})
\b{eqnarray} {\cal
I}(t)&=&\tr[U(t,t_0)\rho(t_0)\ln\rho(t_0)U(t_0,t)]\nonumber\\
&=&\tr[\rho(t_0)\ln\rho(t_0)U(t_0,t)U(t,t_0)]={\cal I}(t_0)\;
,\label{10}\e{eqnarray} because of $U(t_0,t)U(t,t_0)=1$. It is the
information conservation in quantum mechanics\cite{e}.

To measure the entropy of a system, one has to divide the system
into macroscopically infinitesimal parts. The entropy of the $i$th
part is defined to be $S_i=-k_B\tr(\rho_i\ln\rho_i)$, in which
$\rho_i$ is the reduced density operator of the part $i$. The
entropy of the whole system is defined to be the sum \be S=
\sum_iS_i=-k_B\sum_i\tr(\rho_i\ln\rho_i)\label{4}\ee of the
entropies of these parts, as an extensive thermodynamical variable
should be. When one measures the entropy of the system at time
$t_0$, he has destroyed the entanglement of the states of various
parts of the system. The state and the density operator of the
system are therefore factorized. Under this condition, the entropy
of the system is \b{eqnarray}
S(t_0)&=&-k_B\sum_i\tr[\rho_i(t_0)\ln\rho_i(t_0)]\nonumber\\
&=&-k_B\tr[\rho(t_0)\ln\rho(t_0)] =-k_B{\cal I}(t_0)\; .\e{eqnarray}
For an isolated system, the information conservation (\ref{10})
works. The information of the system at $t>t_0$ is therefore \be
{\cal I}(t)=-S(t_0)/k_B\; .\label{s1}\ee During the period from time
$t_0$ to $t$, the interaction between different parts of the system
makes their states be entangled again. It means the states of
different parts are correlated. If one measures the entropy of the
system at time $t$, he has to measure the entropies of every part of
the system, and therefore destroy this entanglement once more. This
is the state collapse, and causes the loss of correlation
information. Since the parts of the system are not isolated, their
information is not conserved. It makes the entropy \be
S(t)=-k_B\sum_i\tr[\rho_i(t)\ln\rho_i(t)]\ee at time $t$ does not
equal $S(t_0)$ in general. By intuition we see, the sum of
information of all parts of the system should not be more than the
information of the system, since the correlation information of
various parts is not included in the sum. It is \be
\sum_i\tr[\rho_i(t)\ln\rho_i(t)]\leq {\cal I}(t)\; .\label{s2} \ee
If this is true, we obtain \be S(t)\geq S(t_0)\label{s3}\ee from
(\ref{s1})-(\ref{s2}) for an isolated system.

To prove the statement (\ref{s2}), let us remind you some
mathematical inequalities.  We also collect the proofs of these
inequalities here, to make our description be self-contained,
although their original forms may be found in text books\cite{e,j}.
By the way, in the following we understand that
$0\ln0\equiv\lim_{\xi\rightarrow 0}(\xi\ln\xi)=0$.

\nt {\bf Lemma 1.} For any non-negative number $x$ we have  \be x\ln
x\geq x-1\; , \label{11}\ee the equality holds when and only when
$x=1$.

Proof: It may be verified by differentiation, that $x\ln x-( x-1)$
as a continuous function of non-negative variable $x$ has unique
minimum 0 at $x=1$. The lemma is therefore proven.

\nt {\bf Lemma 2.} For sets $[w_i]$ and $[x_i]$ of non-negative
numbers with $\sum_ix_i=1$, we have \be
\sum_ix_iw_i\ln\sum_{i^\prime}x_{i^\prime}
w_{i^\prime}\leq\sum_ix_iw_i\ln w_i\; .\label{12}\ee

Proof: The average $\bar{w}\equiv\sum_ix_iw_i$ is non-negative. For
$\bar{w}>0$, by lemma 1 we see
\begin{eqnarray} \sum_ix_iw_i\ln\sum_{i^\prime}x_{i^\prime}w_{i^\prime}-\sum_ix_iw_i\ln
w_i\;\;\;\;\;\;\;\;\;\;\;\;\;\;\;\;\;\;\nonumber\\=-\sum_ix_i\bar{w}\frac{w_i}{\bar{w}}
\ln\frac{w_i} {\bar{w}}\leq
-\sum_ix_i\bar{w}(\frac{w_i}{\bar{w}}-1)=0\;
,\nonumber\end{eqnarray} (\ref{12}) is true. Since two sides of
(\ref{12}) are continuous functions of non-negative variables
$[w_i]$ and $[x_i]$, it is also true for the limit case $\bar{w}=0$.
The lemma is therefore proven.

\nt {\bf Lemma 3.} For sets $[W_i]$ and $[T_{ij}]$ of non-negative
numbers with \be \sum_iW_i=1\;\;\;\;\; \mbox{and}\;\;\;\; \;
\sum_iT_{ij}=\sum_jT_{ij}=1 \; ,\label{0}\ee we have \be W_j^\prime
\equiv\sum_iW_iT_{ij}\geq 0 \;\;\;\;\mbox{for every} \;j
,\label{a}\ee\be\sum_jW_j^\prime=1\; ,\label{b}\ee and \be
\sum_jW_j^\prime\ln W_j^\prime\leq\sum_iW_i\ln W_i\; .\label{13}\ee

Proof: (\ref{a}) and (\ref{b}) are obvious. By (\ref{0}) and lemma 2
we see \b{eqnarray} \sum_jW_j^\prime\ln W_j^\prime &=&\sum_j\left(
\sum_iW_iT_{ij}\right)\ln
\left(\sum_{i^\prime}W_{i^\prime}T_{i^\prime j}\right)\nonumber \\
&\leq & \sum_{ij}W_iT_{ij}\ln W_i=\sum_i W_i\ln W_i\; .\nonumber
\e{eqnarray} The lemma is therefore proven.

\nt {\bf Lemma 4.} For positive numbers $[W_{ij}]$,
$W_i=\sum_jW_{ij}$ and $W_j^\prime=\sum_iW_{ij}$, with
$\sum_{ij}W_{ij}=1$, we have \be
\sum_iW_i=1\;,\;\;\;\;\;\;\sum_jW_j^\prime =1\;,\label{14}\ee and
\be \sum_{ij}W_{ij}\ln W_{ij}\geq \sum_iW_i\ln W_i
+\sum_jW_j^\prime\ln W_j^\prime\; .\label{15}\ee The equality holds
when and only when $W_{ij}=W_iW_j^\prime$ for all $ij$, it is that
the $W_{ij}$ may be factorized.

Proof: (\ref{14}) is obvious. By lemma 1 we see \be
\frac{W_{ij}}{W_iW_j^\prime}\ln \frac{W_{ij}}{W_iW_j^\prime}\geq
\frac{W_{ij}}{W_iW_j^\prime}-1\; ,\label{c} \ee the equality holds
when and only when $W_{ij}=W_iW_j^\prime$. Multiplying two sides of
(\ref{c}) by the positive number $W_iW_j^\prime$ and summing up over
$ij$, one obtains \be \sum_{ij}W_{ij}\ln W_{ij}- \sum_iW_i\ln W_i
-\sum_jW_j^\prime\ln W_j^\prime\geq 0\; .\nonumber\ee This is
exactly (\ref{15}). The lemma is therefore proven.

Suppose $[L]$ is a complete set of commutative dynamical variables
of the system, with a complete orthonormal set of eigenstates
$[|n\rangle]$. The $[L]$ representation of density operator $\rho$
is a matrix with elements $\rho_{n,n^\prime}=\langle
n|\rho|n^\prime\rangle$. If $\rho$ itself is included in the set
$[L]$, the $[L]$ representation of $\rho$ is called natural. In a
natural representation, the density matrix is diagonal:
$\rho_{n,n^\prime}=W_n\delta_{n,n^\prime}$, in which $W_n$ is the
$n$th eigenvalue of $\rho$, denoting the probability of finding the
system being in the state $|n\rangle$. The information (\ref{2}) may
be written in the form \be {\cal I} =\sum_nW_n\ln W_n\;
,\label{d}\ee with a set $[W_n]$ of non-negative numbers. One may
also consider the information about a specially chosen complete set
of commutative dynamical variables $[L^\prime]$, with complete set
of orthonormal eigenstates $[|m\rangle]$. For an ensemble of the
systems with the density operator $\rho$ , the probability of
finding the system in the state $|m\rangle$ is \be
W^\prime_m=\sum_n\langle m|n\rangle W_n\langle n|m\rangle\;
.\label{e}\ee The definition of the information about the variables
$[L^\prime]$ is \be {\cal I}_{[L^\prime]}\equiv\sum_mW^\prime_m\ln
W^\prime_m\; .\ee Since $|\langle n|m\rangle|^2$ are non-negative,
and $\sum_n|\langle n|m\rangle|^2=\sum_m|\langle n|m\rangle|^2=1$,
according to lemma 3 and equation (\ref{d}) we have \be {\cal
I}_{[L^\prime]}\leq {\cal I}\; . \ee

Now, let us divide the system into two parts $a$ and $b$. Suppose
$[L_i]$, with $i=a$ or $b$, is a complete set of commutative
dynamical variables of part $i$, $|n_i\rangle$ is their $n_i$th
eigenstate, and $[|n_i\rangle]$ is a complete set of states of part
$i$. Therefore $[|n_an_b\rangle]\equiv[|n_a\rangle|n_b\rangle]$ is a
complete orthonormal set of states of the system. In the $[L_aL_b]$
representation, The density operator of the system is a matrix, with
elements \be \rho_{n_an_b,n^\prime_an^\prime_b}\equiv\langle
n_an_b|\rho|n^\prime_a n^\prime_b\rangle\; .\ee From (\ref{e}) we
see the probability of finding part $a$ in the state $|n_a\rangle$
and part $b$ in the state $|n_b\rangle$ is \be
W_{n_an_b}=\sum_n\langle n_an_b|n\rangle W_n\langle
n|n_an_b\rangle\; ,\ee with normalization \be
\sum_{n_an_b}W_{n_an_b}=1\; .\label{z}\ee The information of
dynamical variables $[L_a,L_b]$ is \be {\cal
I}_{L_a,L_b}=\sum_{n_a,n_b}W_{n_an_b}\ln W_{n_an_b}\leq {\cal I}\;
.\label{f}\ee The probability of finding part $a$ in the state
$|n_a\rangle$ and the probability of finding the part $b$ in the
state $|n_b\rangle$ are \be
W_{n_a}=\sum_{n_b}W_{n_an_b}\;\;\;\;\mbox{and}\;\;\;\;W^\prime_{n_b}
=\sum_{n_a}W_{n_an_b}\; .\label{g}\ee respectively. In
(\ref{z}-\ref{g}), it is understood that the summation is over those
$n_a$ and $n_b$ only, for which $W_{n_an_b}>0$.

The density operator $\rho_a$ of part $a$ is reduced from the
density operator $\rho$ of the system. In the representation
$[L_a]$, it is a matrix with elements \be
(\rho_a)_{n_a,n^\prime_a}=\sum_{n_b}\rho_{n_an_b,n^\prime_an_b}
=\sum_{n_b}\langle n_an_b|\rho |n^\prime_an_b\rangle\; ,\ee and may
be written in a compact form \be \rho_a=\tr_b\rho \; .\ee The
subscript $b$ denotes that the trace is a sum of matrix elements
diagonal with respect to quantum numbers of part $b$. Likewise,
$\rho_b=\tr_a\rho $. Suppose $\rho_i$ is included in the set
$[L_i]$, the probability of finding the part $i$ in state
$|n_i\rangle$ is its eigenvalue $W_{n_i}$, and is expressed in
(\ref{g}). The information about part $i$ is \be {\cal
I}_i=\tr\rho_i\ln\rho_i=\sum_{n_i}W_{n_i}\ln W_{n_i}\; .\ee From
lemma 4 and equations (\ref{f},\ref{g}) we see \be
\tr\rho_a\ln\rho_a+\tr\rho_b\ln\rho_b\leq\tr\rho\ln\rho\;
.\label{h}\ee The equality holds when and only when the density
operator of the system may be factorized into a direct product of
density operators of its parts, it is when and only when its parts
do not correlate with each other. We may further subdivide the parts
and apply (\ref{h}) to them again and again, the result is the
statement (\ref{s2}). As we showed before, this statement proves
(\ref{s3}), which is the\vskip 0.1cm

\nt {\bf Theorem:} The entropy of an isolated system if changes can
only increase. \vskip 0.1cm

\nt It is exactly the second law of thermodynamics. This law is
therefore finally proven. According to the relationship between the
entropy and the probability of a macroscopic state\cite{l} referred
at the beginning of this paper, it in turn shows that an isolated
macroscopic system always goes from the less probable state to the
more probable state.

The proof here is quite general. It looks like relying on the
quantum mechanical effects of state entanglement and its collapse.
However, it is still more general. It is an information theoretical
proof, relies only on the conservation (\ref{10}) and the
non-additivity (\ref{h}) of the information. The extensive character
(\ref{4}) (additivity) of the entropy is also important. Information
conservation is a character of dynamics. It is shared by quantum
dynamics and classical dynamics, as well as some possible dynamics
not yet have been discovered. The non-additivity of information is
purely mathematical. It may be deduced from the general relations
(\ref{g}) of the probabilities by use of mathematical inequalities
stated before. State entanglement and its collapse is only a special
way of their realization. They may be realized in classical
mechanics or in some unknown mechanics as well. The second law of
thermodynamics is therefore almost dynamics independent, except the
requirement of information conservation. It may be still exactly
true in the future, even though one day people find that the quantum
mechanics is only approximate. It is also quite generally
applicable, not only to thermodynamics but also to some other
statistical sciences, if the information conservation is true for
them. To consider its possible applications in the social and
financial sciences is interesting.

From the proof we learn that the entropy of an isolated system
increases only because one loses the correlation information between
different parts of the system. It emphasizes the importance of the
correlation information in a complete statistical science.

This work is supported by the National Nature Science Foundation of
China with Grant number 10305001.

\b{\t}{99} \bi{l} See for example L. D. Landau and E. M. Lifshitz
Statistical Physics, 3rd edition (Butterworth Heinemann, Oxford,
1980) \bi{s1}C.E. Shannon, B.S.T.J. {\bf 27}(1948) 379, 626
\bi{s2}C.E. Shannon and W Weaver, The mathematical theory of
communication (University of Illinois press, Urbana, I11., 1949)
\bi{e}H. Everett, The theory of universal wave function in the
many-worlds interpretation of quantum mechanics, B.S.De Witt and N.
Graham eds. (Princeton University press, Princeton, 1973)
\bi{j}E.T.Jaynes, Probability theory: the logic of science
(Cambridge University press, Cambridge, 2003)\e{\t}
\end{document}